\documentstyle[twocolumn,aps]{revtex}
\begin{document}
\title{On the Fermionic Quasi-particle Interpretation in Minimal Models of
Conformal Field Theory}
\author{{\bf \large A.A. Belavin $^{\dag}$ and A. Fring $^*$}}
\address{ \noindent $^{\dag}$ Landau Institute for Theoretical Physics
,Chernogolovka,
142432, Russia \\
$*$ Institut f\"ur Theoretische Physik,
Freie Universit\"at Berlin, Arnimallee 14, D-14195 Berlin, Germany}
\date{\today}
%\thanks{
%Belavin@laudau.itp.ru \newline
%Fring@physik.fu-berlin.de}
\maketitle

\begin{abstract}
The conjecture that the states  of the ferminonic
quasi-particles in minimal conformal field theories are eigenstates of the
integrals of motion to certain eigenvalues is checked and shown to be
correct only for the Ising model.
\end{abstract}

%******************************************************************

\section{Introduction}

The Hilbert spaces of two dimensional conformal field theories \cite{BPZ}
are described in terms of chiral algebras which act on a finite set of
fields. The basis of these spaces are however not unique, which in turn, on
the basis of the characters, leads to very interesting generalizations of
the famous identities of Rogers-Schur and Ramanujan \cite{RSR,Hardy}
appearing in number theory.

Character formulae for irreducible highest weight representations of the
Virasoro algebra ({\it {Vir}}) exist in several alternative forms. The
oldest, very often refered to as bosonic representation, are the formulae of
Feigin and Fuchs and Rocha-Caridi \cite{FeiginFuchs,Rocha}, which directly
incorporate the structure of the Null-vectors, i.e. divides out the invariant
ideal. When interpreted as partition function one demands modular invariance
of these expressions. In order to investigate this property it is most
natural to re-express the characters in terms of Theta functions for which
these transformations are well-known. Furthermore via the Jacobi triple
identity it is possible to establish an easy relation to the currently
ubiquitous quantum dilogarithm \cite{Faddeev}, which is very useful to carry
out semi-classical limits. In order to understand the relation to massive
integrable theories, or more precisely perturbed conformal field theories 
\cite{Zamo1} more recent formulae, also known as fermionic representations
are most suitable. These expressions posses a remarkable direct fermionic
interpretation in terms of quasi-particles for the states, obeying Pauli's
exclusion principle. The connection between these states and the spectrum of
quasi-particle excitations, which arise from the Bethe Ansatz equations for
the eigenvalues of the Hamiltonians, have recently been elaborated by the
Stony Brook group \cite{McCoy,KKMM1,KKMM2,McCoy2} .

Whereas the former bosonic representations are in general unique, the latter
are not, possibly indicating different relevant perturbations already at the
conformal point of the theory. Ultimately one would like to construct these
states explicitly in terms of cosets \cite{FeiginSto}, in a sense we shall
specify below.

In the present letter we will check the simple conjecture that the fermionic
states are eigenvalues of the integrals of motion of the same form as in the
continuum theory.

\section{The quasi-particle interpretation}

For the convenience of the reader and in order to establish our notation we
shall briefly recall various equivalent expressions for the characters and
then explain how they lead to an interpretation in terms of quasi-particles.
First of all the bosonic representation \cite{FeiginFuchs,Rocha} for the
minimal models ${\cal M}(s,t)$ \cite{BPZ} with central charge $c=1-\frac{%
6(s-t)^{2}}{s\;t}$ and highest weight $h_{n,m}=\frac{(ns-mt)^{2}-(s-t)^{2}}{%
4\;s\;t}$ reads 
\begin{equation}
\chi _{n,m}^{s,t}(q)=\frac{q^{-\frac{c}{24}}}{(q)_{\infty }}\sum_{k=-\infty
}^{\infty }\left( q^{h_{n+2kt,m}}-q^{h_{n+2kt,-m}}\right) \;\;\;.
\end{equation}
Here we have used the standard abbreviation $(q)_{m}=\prod%
\limits_{k=1}^{m}(1-q^{k})$. Introducing the parameterization $q=e^{i\pi
\tau }$, $k_{\pm }=\tau \pi (s\;n\pm m\;t)$, $k_{0}=\tau s\;t$ and using the
well known formula for one of the Theta functions $\Theta _{3}(p,\tau
)=\sum\limits_{k=-\infty }^{+\infty }q^{k^{2}}e^{ipk}$ one easily derives 
\[
\chi _{n,m}^{s,t}(q)=\frac{1}{\eta (q)}\left( q^{\frac{k_{-}^{2}}{%
4\;k_{0}\tau \pi ^{2}}}\Theta _{3}(k_{-},k_{0})-q^{\frac{k_{+}^{2}}{%
4\;k_{0}\tau \pi ^{2}}}\Theta _{3}(k_{+},k_{0})\right) 
\]
with $\eta (q)$ denoting Dedekind's Eta function. The relation to quantum
dilogarithms \cite{Faddeev} $S_{q}(p)=\prod\limits_{l=1}^{\infty
}(1+e^{ip}q^{2l-1})$ is easily established by employing Jacobi's triple
identity 
\begin{equation}
\Theta _{3}(q,\tau )\;=\;S_{q}(p)\;S_{q}(-p)\;S_{q}(\pi \tau )\;\;.
\end{equation}
Now one is in a nice position to carry out semi-classical limits, by knowing
that the quantum dilogarithm acquires classically the form of Roger's
dilogarithm \cite{Lewin} 
\begin{equation}
\lim\limits_{\tau \rightarrow 0}S_{q}(\pi +p)\;=\;\exp \left( \frac{1}{2\pi
i\tau }\;L_{2}(e^{ip})\;+\;{\cal O}(\tau )\right) \;\;\;.
\end{equation}
For the fermionic representation there exist two versions, which are of
slightly different nature when interpreted as partition functions. First 
\begin{equation}
\chi _{A,\vec{B}}^{\vec{m}}(q)=\sum_{\vec{m}=0}^{\infty }\frac{q^{\vec{m}A%
\vec{m}^{t}+\vec{m}\cdot \vec{B}}}{(q)_{m_{1}}\ldots (q)_{m_{n}}}
\label{eq: cf1}
\end{equation}
where $A$ is a  $N \times N$-matrix, with being the number of species,
and $\vec{B}$ denotes a vector which needs to be
specified for a particular theory and super-selection sector. The summation
over $m_{1},m_{2},\ldots $ may be restricted in some way indicating that
certain particles may only appear in conjunction with others. Choosing $A$
to be the inverse of the Cartan matrix $C$, related to a particular simply
laced Lie algebra, it was found in \cite{KKMM1} that for $A_{n}$ one obtains
the $Z_{n+1}$ invariant parafermionic theories, for the $D_{n}$ case the $r=%
\sqrt{n/2}$ orbifold theories and for $E_{6},E_{7},E_{8}$ the tricritical
three state Potts -, the tricritical Ising - and the Ising model,
respectively. Different restrictions on the summation correspond in this
case in general to different symmetries of the Dynkin diagram and different
vectors $\vec{B}$ are related to different super-selection sectors. In \cite
{KKMM1} these expressions where found by means of Mathematica and to the
knowledge of the authors explicit analytic proofs are still lacking for most
of them. An exception is the character formula for the $h_{11}=0$ sector of
the $E_{8}$-theory, for which Warnaar and Pearce \cite{WP} utilized the
connection between the Ising model and the dilute $A_{3}$-model in order to
proof it.

Of different nature are the expressions for the minimal models 
${\cal M}(s=l+2,t=l+3)$ which are entirely based on $A_{l}$ 
\begin{eqnarray}
\chi _{n,m}^{l}(q) &=&\sum_{\vec{k}}\frac{q^{\frac{1}{4}\vec{k}C\vec{k}^{t}-%
\frac{1}{2}\vec{e}_{l+2-m}\cdot \vec{k}-\frac{1}{4}(m-n)(m-n-1)}}{(q)_{k_{1}}%
}  \nonumber \\
&&\hspace{0.8cm}\prod_{a=2}^{l}\left[ 
%TCIMACRO{
%\QATOP{{\frac{1}{2}\left( \vec{k}I_{l}-\vec{e}_{n}+\vec{e}_{l+2-m}\right) _{a}}}{{k_{a}}} }
%BeginExpansion
{{\frac{1}{2}\left( \vec{k}I_{l}-\vec{e}_{n}+\vec{e}_{l+2-m}\right) _{a}} \atop {k_{a}}}%
%EndExpansion
\right] _{q}  \label{eq: cf2}
\end{eqnarray}
where the q deformed binomial coefficient 
\begin{equation}
\left[ 
%TCIMACRO{\QATOP{n}{m} }
%BeginExpansion
{n \atop m}%
%EndExpansion
\right] _{q}=\frac{(q)_{n}}{(q)_{m}(q)_{n-m}}
\end{equation}
has been introduced and the sum is restricted to $\vec{k}\in
(2Z\!\!\!Z)^{n}+(m-1)(\vec{e}_{1}+\ldots \vec{e}_{l})+(\vec{e}_{n-1}+\vec{e}%
_{n-3}\ldots )+(\vec{e}_{l+3-m}+\vec{e}_{l+5-m}\ldots )$ .
Furthermore $\vec{e}_{j}$ denotes the $l$-dimensional unit vector in the
direction $j$ and $I_{l}$ is the incidence matrix $I_{l}=2-C_{l}$, 
of $A_{l}$. This formula was proven by Berkovich et. al. \cite{Berk}.
(\ref{eq: cf1}) and (\ref{eq: cf2}) only coincide for $c=\frac{1}{2},\frac{4%
}{5}$. As shown by Kedem, Klassen, McCoy and Melzer both possess an
interpretation in terms of fermionic
quasi-particles. For this purpose one considers the characters as partition
function $Z$ 
\begin{equation}
\chi \sim Z=\sum_{{\rm states}}e^{-\frac{{\rm E(states)}}{kT}%
}=\sum_{l=0}^{\infty }P(E_{l})e^{-\frac{E_{l}}{kT}}
\end{equation}
with T being as usual the Temperature, $k$ Boltzmann's constant, $E_{l}$
and $P(E_{l})$ the energy and the degeneracy of the particular level $l$,
respectively. Regarding the system as a gas of particles in a box of size L,
L is thought to be large, one may quantize the possible momenta in units of $%
2\pi /L$. Assuming that there are n different species of particles one
obtains this way a set of single particle momenta $p_{i_{a}}^{a}$ for the
quasi-particles, characterized by $a$, the particle type and a non-negative
integer $i_{a}$, which is subject to some restrictions depending on the
particular model. The energy spectrum minus the ground state energy and the
one for the related momenta may then be expressed as 
\begin{equation}
E_{l}\;=\;\sum_{a=1}^{n}\sum_{i_{a}=1}^{m_{a}}\;e_{a}(p_{i_{a}}^{a})\qquad
p_{l}\;=\;\sum_{a=1}^{n}\sum_{i_{a}=1}^{m_{a}}\;p_{i_{a}}^{a}.
\label{eq: quasp}
\end{equation}
By definition if a many-body system obeys (\ref{eq: quasp}) in the infinite
size limit, its spectrum is said to be of quasi-particle type. For the
quasi-particles to be of fermionic nature one requires that one of the
restrictions acquires the form of Pauli's exclusion principle 
\begin{equation}
p_{i_{a}}^{a}\;\neq \;p_{j_{a}}^{a}\qquad \hbox{for}\;{i_{a}\neq j_{a}}\;.
\end{equation}
Formally this is achieved by employing the well known formula from number
theory, (refer for instance \cite{Hardy}) for the number of partitions $%
P_{M}(n,m)$ of a non-negative integer n into M distinct non-negative
integers which are smaller than m 
\begin{equation}
\sum_{n=0}^{\infty }P_{M}(n,m)q^{n}\;=\;q^{\frac{1}{2}M(M-1)}\left[ 
%TCIMACRO{\QATOP{m+1}{M} }
%BeginExpansion
{m+1 \atop M}%
%EndExpansion
\right] _{q}\;\;\;.  \label{party}
\end{equation}
In case there is no upper limit, i.e. when m tends to infinity, we simply
employ on the right hand side 
\begin{equation}
\lim_{m\rightarrow \infty }\left[ 
%TCIMACRO{\QATOP{m+1}{M} }
%BeginExpansion
{m+1 \atop M}%
%EndExpansion
\right] _{q}\;=\;\frac{1}{(q)_{M}}\;\;.
\end{equation}
The requirement of distinctiveness expresses here the fermionic nature of
the quasi-particles in this Ansatz. Employing (\ref{party}) in (\ref{eq: cf1}%
) and choosing $q=e^{-\frac{2\pi v}{kTL}}$, $v$ being the velocity,
it is straightforward to derive
that the possible set of momenta is 
\begin{equation}
p_{i_{a}}^{a}\;=\;\frac{2\pi }{L}\left( \left( A_{ab}-\frac{1}{2}\right)
m_{b}+B_{a}+\frac{1}{2}+N_{i_{a}}\right)
\end{equation}
where $N_{i_{a}}$ is a set of non-negative distinct integers. Proceeding in
the same way for (\ref{eq: cf2}) one ends up with some restrictions from
above for the possible momenta for all particles except the first one. This
feature distinguishes (\ref{eq: cf2}) from (\ref{eq: cf1}), since in the
latter case the particles are more on the same footing, whereas in the
former case particle one plays the dominant role. One obtains for the
possible momenta, in units of $\frac{2\pi }{L}$, of the minimal models 
\begin{eqnarray*}
p_{i_{1}}^{1} &\in &\left\{ p_{1}^{{\em min}}(\vec{k}),p_{1}^{{\em min}}(%
\vec{k})+1,p_{1}^{{\em min}}(\vec{k})+2,\ldots \right\} \\
p_{i_{a}}^{a} &\in &\left\{ p_{a}^{{\em min}}(\vec{k}),p_{a}^{{\em min}}(%
\vec{k})+1,\ldots ,p_{a}^{{\em max}}(\vec{k})\right\} \;\;\;
\end{eqnarray*}
with 
\begin{eqnarray*}
\vec{p}^{{\em min}}(\vec{k}) &=&\frac{1}{2}\left( \vec{e}_{1}+\vec{e}%
_{2}+\ldots +\vec{e}_{l}-\vec{e}_{l+2-m}-\frac{\vec{k}\cdot I_{l}}{2}\right)
\\
\vec{p}^{{\em max}}(\vec{k}) &=&\frac{1}{2}\left( \vec{e}_{n}-\vec{e}_{1}-%
\vec{e}_{2}-\ldots -\vec{e}_{l}+\frac{\vec{k}\cdot I_{l}}{2}\right)
\end{eqnarray*}
In this way one may associate to each energy level some well defined set of
fermionic quasi-particle momenta 
\begin{equation}
|p_{1}^{1},\ldots ,p_{1}^{m_{1}},\ldots ,p_{n}^{1},\ldots
,p_{n}^{m_{n}}\rangle \;\;,
\end{equation}
which are in one-to-one correspondence to the decomposition of the Hilbert
space 
\begin{equation}
{\cal H}_{h_{n,m}}\;=\bigoplus\limits_{l=0}^{\infty }\;|h_{n,m}^{l}\rangle
\end{equation}
in form of the irreducible representations of the Virasoro algebra 
\begin{eqnarray*}
L_{0}|h_{n,m}\rangle &=&h_{n,m}|h_{n,m}\rangle \\
L_{k}|h_{n,m}\rangle &=&0\qquad \hbox{for }\;\;\;k>0 \\
|h_{n,m}^{l}\rangle &=&L_{-n_{1}}L_{-n_{2}}\ldots L_{-n_{n}}|h_{n,m}\rangle
\qquad \hbox{for }\;\;\;n_{i}>0\;\;.
\end{eqnarray*}
Here $l=n_{1}+\ldots +n_{n}$ denotes the $l^{{\rm th}}$ level of the
irreducible highest weight module with respect to the weight $h_{n,m}$. The
question of how to construct these states explicitly in terms of cosets of
the Kac-Moody algebra arises naturally, but is still an open problem \cite
{FeiginSto}. Here we shall be less ambitious and only try to find some
further properties of these quasi-particle states. An explicit construction
will of course ultimately also allow to answer this question. For algebras
based on $\hat{sl}(2)_{k}$ and $\hat{su}(k)_{1}$ such a constructions have
been provided in terms of a spinon basis \cite{Bou}. 
% ***********************************************************************

\section{Integrals of Motion}

In the case of massive integrable models conserved charges serve as a very
powerful tool. By assuming locality in the momentum space and the
possibility of diagonalising them as one-particle asymptotic states, they
may be employed, even without knowing their explicit form, to construct the
scattering matrix of a purely elastic scattering theory
\cite{Zamo1}. For the massless models such infinite Abelian subalgebra of
integrals of motions are known to exist in the enveloping algebra of the
Virasoro algebra ({\it UVir}) \cite{Sas}.
The charge densities $T_{2k}(z)$ acquire unique expressions from the
requirement of mutual commutativity and the assignment of a definite spin.
For a chiral field in the plane this may be achieved algebraically by
demanding the condition for a primary field $T_{2k}(z)$ with conformal
dimension $2k$, $(k\in I\!\!N)$ 
\begin{equation}
\left[ L_{n},T_{2k}(z)\right] \;=\;z^{n+1}\frac{dT_{2k}(z)}{dz}%
\;+\;2k\;(n+1)\;z^{n}T_{2k}(z)
\end{equation}
only to be satisfied for the M\"{o}bius subalgebra, i.e. for $n=\pm 1,0$.
This means $T_{2k}(z)$ is asked to be a quasi-primary field. Regarding the
map from the plane to the cylinder as a particular conformal transformation,
i.e. $z=e^{\omega }$, one has a well defined procedure to obtain the spin
densities defined on the cylinder, for instance $T_{2}^{{\em cyl.}}(\omega
)=z^{2}T(z)-\frac{c}{24}$. Integration of the charge densities 
will give the integrals of motion of spin value $2k-1$ 
\begin{equation}
I_{2k-1}\;=\;\int_{0}^{2\pi }\frac{d\omega }{2\pi }T_{2k}(\omega )\;.
\end{equation}
The first of them read \cite{Sas,BLZ} 
\begin{eqnarray}
I_{3} &=&2\sum_{n=1}^{\infty }L_{-n}L_{n}\;+\;L_{0}^{2}\;-\;\frac{2+c}{12}%
L_{0}+k_{3}  \label{eq: int3} \\
I_{5} &=&\sum_{n,m,l}:L_{n}L_{m}L_{l}:\delta _{n+m+l,0}+\frac{3}{2}%
\sum_{n=1}^{\infty }L_{1-2n}L_{2n-1}  \nonumber \\
&&+\sum_{n=1}^{\infty }\left( \frac{11+c}{6}n^{2}-\frac{c}{4}-1\right)
L_{-n}L_{n}   \label{eq: int5} \\
&&-\frac{4+c}{8}L_{0}^{2}+\frac{(2+c)(20+3c)}{24^{2}}L_{0}+k_{5}\;\;. 
\nonumber
\end{eqnarray}
Here $k_{3},k_{5}$ are constants which depend on the super-selection sector.
It will turn out that for our purposes these constants have to be chosen
differently than in \cite{BLZ}.

$:\hspace{0.2cm}:$ is the usual normal ordering prescription which arranges
the operators $L_{n}$ into an increasing sequence with respect to their mode
index. We shall now employ these charges in order to give some
characterization of the quasi-particle states. On the representation space
of the Virasoro algebra they possess a well-defined action and one may
compute explicitly their eigenvalues for each level 
\begin{equation}
I_{s}|h_{n,m}^{l}\rangle \;=\;\lambda _{n,m}^{(s)}|h_{n,m}^{l}\rangle \;.
\end{equation}
Denoting by ${\cal P}_{h_{n,m}^{l}}$ the particular set of momenta which
correspond to the level $l$ in the Verma module of $h_{n,m}^{l}$ the
conjecture that the quasi-particle states are eigenstates of the integrals
of motion arises naturally 
\begin{equation}
I_{s}\,{\cal P}\;=\;\gamma _{{\cal P}_{h_{n,m}^{l}}}^{(s)}{\cal P}\;\;,\;
\end{equation}
with ${\cal P=}|p_{1}^{1},\ldots ,p_{1}^{m_{1}},\ldots ,p_{n}^{1},\ldots
,p_{n}^{m_{n}}\rangle $. Drawing an analogy to the continuum theory, where
the conserved charges act additively and diagonal on asymptotic one-particle
states one may conjecture for the eigenvalues 
\begin{equation}
\lambda _{h_{n,m}^{l}}^{(s)}\;=\;\gamma _{{\cal P}_{h_{n,m}^{l}}}^{(s)}=%
\sum_{a=1}^{N}\chi _{a}^{(s)}\sum_{j_{a}=1}^{m_{a}}\left(
p_{a}^{j_{a}}\right) ^{s}\;+\;\hbox{const}\;\;.  \label{Prop}
\end{equation}
% ********************************************

\subsection{The Ising Model}

We shall start by verifying this conjecture for the Ising Model, i.e. $c=%
\frac{1}{2}$, for which we expect it certainly to be true since it is known
to be equivalent to a free fermion theory. In this case we have the
following identities for the characters 
\begin{eqnarray}
q^{\frac{1}{48}}\chi _{1,1}^{3,4}(q) &=&\sum_{m=0}^{\infty }\frac{%
q^{2\;m^{2}}}{(q)_{2m}} \\
q^{\frac{1}{48}}\chi _{1,3}^{3,4}(q) &=&\sum_{m=0}^{\infty }\frac{%
q^{2\;m^{2}+2m+\frac{1}{2}}}{(q)_{2m+1}} \\
q^{\frac{1}{48}}\chi _{1,2}^{3,4}(q) &=&\sum_{m=0}^{\infty }\frac{q^{\frac{%
m^{2}+m}{2}-\frac{1}{16}}}{(q)_{m}}
\end{eqnarray}
The results for the eigenvalue computation for the highest weight
representation $h_{11}$ are given in table I and II. The eigenvalues $\lambda 
$ are computed explicitly by acting with the integrals of motion on the
highest weight representation. We present the computation until level nine,
since then the second Null-vector appears in the Verma modul. For the
computation it is not necessary  to
divide out the complete invariant ideal, but only all descendants of 
$L_{-1}$, which is a Null-vector at level zero. 
This way one obtains of course more eigenvalues, but one may easily match
them with the values of the appropriate $\gamma $. In principle the
constants $\chi _{1}^{(s)}$ could have been fixed already after the first
few levels and all higher levels serve as a consistency check.
The constants $k_{3},k_{5}$ were found to be different in the Ramond- and
Neveu-Schwarz sector 
\begin{equation}
k_{3}=\frac{55}{3\cdot 16^{2}}\;\delta _{\frac{1}{16},h}\qquad k_{5}=\frac{%
2161}{9\cdot 16^{3}}\;\delta _{\frac{1}{16},h}\quad .
\end{equation}
For the constants $\chi $ we obtain 
\begin{equation}
\chi _{1}^{(3)}\;=\;\frac{7}{6}\qquad \chi _{1}^{(5)}\;=\;\frac{143}{144}%
\;\;.
\end{equation}
Notice that the $\chi _{1}^{(s)}$ do not depend on h, i.e. they are
universal constants independent of the super-selection sector.
\subsection{The unitary minimal Models}

We will now carry out a similar argumentation for the unitary minimal
models. First we consider the $11$-sector, for which (\ref{eq: cf2}) takes
on a particular simple form 
\begin{equation}
\chi_{1,1}^l(q) = \sum_{\vec{k}\in ( 2 Z \!\!\! Z)^2 } \frac{ q^{\frac{1}{4} 
\vec{k} \cdot C \cdot \vec{k}^t }}{(q)_{l_1} } \prod_{a=2}^{l} \left[ 
%TCIMACRO{
%\QATOP{{\frac{1}{2} \left( \vec{k} I_l - \vec{e}_{1} \right)_a} }{{\ k_a } } }
%BeginExpansion
{{\frac{1}{2} \left( \vec{k} I_l - \vec{e}_{1} \right)_a}  \atop {\ k_a } }%
%EndExpansion
\right]_q
\end{equation}
such that the minimal momentum becomes 
\newpage

\begin{table}[h]
\begin{tabular}{l|l|l|l}
l & {\rm d} & $\lambda^{(3)}_{h_{1,1}^l }$ & $\gamma^{(3)}_{{\cal P}%
_{h_{1,1}^l} }$ \\ \hline
0 & 1 & 0 & 0 \\ 
1 & 0 & - & - \\ 
2 & 1 & $\frac{49}{12}$ & $\chi_1^{(3)} \frac{7}{2}$ \\ 
3 & 1 & $\frac{147}{8}$ & $\chi_1^{(3)} \frac{63}{4}$ \\ 
4 & 2 & ($\frac{133}{6},\frac{301}{6}) $ & $\chi_1^{(3)} (43,19)$ \\ 
5 & 2 & ($\frac{1295}{24},\frac{2555}{24}) $ & $\chi_1^{(3)} (\frac{365}{4},%
\frac{185}{4})$ \\ 
6 & 3 & ($\frac{273}{4},\frac{441}{4},\frac{777}{4} ) $ & $\chi_1^{(3)} (%
\frac{333}{2},\frac{189}{2},\frac{117}{2} )$ \\ 
7 & 3 & ($\frac{2989}{24},\frac{4753}{24},\frac{7693}{24} ) $ & $%
\chi_1^{(3)} (\frac{1099}{4},\frac{679}{4},\frac{427}{4} )$ \\ 
8 & 5 & ($\frac{217}{3},\frac{469}{3},\frac{637}{3}, \frac{973}{3},\frac{1477%
}{3} ) $ & $\chi_1^{(3)} (422,278,182,134,62)$ \\ 
9 & 5 & ($\frac{1029}{8},\frac{1953}{8},\frac{2709}{8}, \frac{3969}{8},\frac{%
5733}{8} ) $ & $\chi_1^{(3)} (\frac{2457}{4},\frac{1701}{4},\frac{1161}{4} ,%
\frac{837}{4},\frac{441}{4} )$ \\ 
\end{tabular}
\caption{Eigenvalues of $I_3$ for the $h=0$ sector}
\end{table}

\begin{table}[h]
\begin{tabular}{l|l|l|l}
l & {\rm d} & $\lambda^{(5)}_{h_{1,1}^l }$ & $\gamma^{(5)}_{{\cal P}%
_{h_{1,1}^l} }$ \\ \hline
0 & 1 & 0 & 0 \\ 
1 & 0 & - & - \\ 
2 & 1 & $\frac{8723}{1152}$ & $\chi_1^{(5)} \frac{61}{8}$ \\ 
3 & 1 & $\frac{74503}{768}$ & $\chi_1^{(5)} \frac{1563}{16}$ \\ 
4 & 2 & ( $\frac{60203}{24^2}$, $\frac{300443}{24^2}$ ) & $\chi_1^{(5)}( 
\frac{421}{4}, \frac{2101}{4} )$ \\ 
\end{tabular}
\caption{Eigenvalues of $I_5$ for the $h=0$ sector}
\end{table}

Table III and IV show some values for the equivalent computation for the other
super-selection sectors.

\begin{table}[h]
\begin{tabular}{l|l|l|l|l|l}
l & {\rm d} & $\lambda^{(3)}_{h_{2,1}^l }$ & $\gamma^{(3)}_{{\cal P}%
_{h_{2,1}^l} }$ & $\lambda^{(5)}_{h_{2,1}^l }$ & $\gamma^{(5)}_{{\cal P}%
_{h_{2,1}^l} }$ \\ \hline
0 & 1 & $\frac{7}{48}$ & $\chi_1^{(3)} \frac{1}{8}$ & $\frac{143}{4608}$ & $%
\chi_1^{(5)} \frac{1}{32}$ \\ 
1 & 1 & $\frac{63}{16}$ & $\chi_1^{(3)} \frac{27}{8}$ & $\frac{3861}{512}$ & 
$\chi_1^{(5)} \frac{243}{32}$ \\ 
2 & 1 & $\frac{875}{48}$ & $\chi_1^{(3)} \frac{125}{8}$ & $\frac{446875}{4608%
}$ & $\chi_1^{(5)} \frac{3125}{32}$ \\ 
\end{tabular}
\caption{Eigenvalues of $I_3,I_5$ for the $h=\frac{1}{2}$ sector}
\end{table}
\begin{table}[h]
\begin{tabular}{l|l|l|l|l|l}
l & {\rm d} & $\lambda^{(3)}_{h_{1,2}^l }$ & $\gamma^{(3)}_{{\cal P}%
_{h_{1,2}^l} }$ & $\lambda^{(5)}_{h_{1,2}^l }$ & $\gamma^{(5)}_{{\cal P}%
_{h_{1,2}^l} }$ \\ \hline
0 & 1 & $\frac{1}{16}$ & $\frac{1}{16}$ & $\frac{1}{16}$ & $\frac{1}{16}$ \\ 
1 & 1 & $\frac{59}{48}$ & $\chi_1^{(3)} + \frac{1}{16} $ & $\frac{19}{18}$ & 
$\chi_1^{(5)} + \frac{1}{16} $ \\ 
2 & 1 & $\frac{451}{48}$ & $\chi_1^{(3)}\; 8+ \frac{1}{16}$ & $\frac{4585}{%
144}$ & $\chi_1^{(5)} 32 + \frac{1}{16} $ \\ 
\end{tabular}
\caption{Eigenvalues of $I_3,I_5$ for the $h=\frac{1}{16}$ sector}
\end{table}
\noindent 
\begin{table}[h]
\begin{tabular}{l|l|l|l}
level & d & momentum 1 & momentum 2 \\ \hline\hline
1 & 0 & -- & -- \\ \hline
2 & 1 & $p_1^1(2,0,0,\ldots) = \frac{1}{2} $ & $p_2^1(2,0,0,\ldots) = \frac{3%
}{2} $ \\ \hline
3 & 1 & $p_1^1(2,0,0,\ldots) = \frac{1}{2} $ & $p_2^1(2,0,0,\ldots) = \frac{5%
}{2} $ \\ \hline
4 & 2 & $p_1^1(2,0,0,\ldots) = \frac{1}{2} $ & $p_2^1(2,0,0,\ldots) = \frac{7%
}{2} $ \\ 
&  & $p_1^1(2,0,0,\ldots) = \frac{3}{2} $ & $p_2^1(2,0,0,\ldots) = \frac{5}{2%
} $%
\end{tabular}
\caption{Fermionic states in the unitary minimal models}
\end{table}

%**********************************************************************
\begin{equation}
\vec{p}^{{\em \;\;min}}(k_1,k_2,\ldots,k_l) = \frac{1}{2} \left( 
\begin{tabular}{c}
1 \\ 
1 \\ 
1 \\ 
1 \\ 
\vdots
\end{tabular}
\right) - \frac{1}{4} \left( 
\begin{tabular}{c}
$k_2$ \\ 
$k_1 + k_3$ \\ 
$k_2 + k_4$ \\ 
$k_3 + k_5 $ \\ 
\vdots
\end{tabular}
\right)
\end{equation}

This means in \underline{all} unitary minimal models the structure of the
first levels may be build up entirely from particle one alone. Compare 
table V for this.

To be more general we keep now in (\ref{eq: int3}) also the variable in
front of $L_{0}$ to be arbitrary, say $k_{3}^{0}$. By the same procedure as
in the previous section the action of $I_{3}$ on these states leads to the
following equations 
\begin{eqnarray*}
4+c+2\;k_{3}^{0}+k_{3} &=&\frac{7}{2}\chi _{1}^{(3)} \\
17+c+3\;k_{3}^{0}+k_{3} &=&\frac{63}{4}\chi _{1}^{(3)} \\
34+6\;c+\sqrt{148+88c+16c^{2}}+4\;k_{3}^{0}+k_{3} &=&43\chi _{1}^{(3)} \\
34+6\;c-\sqrt{148+88c+16c^{2}}+4\;k_{3}^{0}+k_{3} &=&19\chi _{1}^{(3)} \;\; .
\end{eqnarray*}
It turns out that this is in fact the only solution for these equations and
hence (\ref{Prop}) only holds for free fermions. In principle the assignment
of the right hand side to the left hand side in the last two equations might
have been reversed, but we expect to recover the Ising model as a particular
case and therefore we have chosen the above relations. It is remarkable that
when taking the values of (\ref{eq: int3}) up to level 3 the above equation
become true identities for all values of the central charge. Proceeding in
the same way for (\ref{eq: int5}), naming the constant in front of $L_{0}$ $,
k_{5}^{0}$, leads to the following equations 
\begin{eqnarray}
\frac{91}{12}-\frac{19}{48}c+2\;k_{5}^{0}+k_{5} &=&\frac{61}{8}\chi
_{1}^{(5)}  \nonumber \\
\frac{581}{6}-\frac{5}{24}c+3\;k_{5}^{0}+k_{5} &=&\frac{1563}{16}\chi
_{1}^{(5)}  \nonumber
\end{eqnarray}
Choosing $k_{5}$ to be zero in order to be able 
to recover the Ising model we obtain
already from these two equations that in fact $c=\frac{1}{2}$ is the only
solution. Hence the fact that for the $I_{3}$ integral of motion the third
level allowed an arbitrary solution must be viewed as a coincidence.

\section{Conclusions}
Similarly as in the last section we may proceed for the other minimal models.
For instance we have carried out the equivalent computation for the 
${\cal M}(2,5)$ (the Yang-Lee edge ) model, also with the same negative 
answer. In conclusion we can say that, except for the Ising model, the 
integrals of motion do not act in the same way on the fermionic 
quasi-particle states as expected from the continuum theory. It would be very
interesting to investigate whether these states are at all eigenstates of 
the integrals of motion.

{\bf Acknowledgment: } A.B. is grateful to the members of the Institut
f\"ur Theoretische Physik at the Freie Universit\"at Berlin, especially
to R. Schrader and M. Karowski, for discussions and warm hospitality. 
A.F. would like to thank the Landau Institut for their
kind hospitality.

%*******************************************************************

\end{document}